\documentclass[sigconf]{acmart}

\AtBeginDocument{%
  \providecommand\BibTeX{{%
    \normalfont B\kern-0.5em{\scshape i\kern-0.25em b}\kern-0.8em\TeX}}}

\usepackage{multirow}

\usepackage{enumitem}
\setlist[itemize]{leftmargin=*}
\setcopyright{acmcopyright}

\begin{document}
\fancyhead{}

\copyrightyear{2020}
\acmYear{2020}
\setcopyright{acmcopyright}\acmConference[ICMR '20]{Proceedings of the 2020 International Conference on Multimedia Retrieval}{June 8--11, 2020}{Dublin, Ireland}
\acmBooktitle{Proceedings of the 2020 International Conference on Multimedia Retrieval (ICMR '20), June 8--11, 2020, Dublin, Ireland}
\acmPrice{15.00}
\acmDOI{10.1145/3372278.3390677}
\acmISBN{978-1-4503-7087-5/20/06}

\title{Knowledge Enhanced Neural Fashion Trend Forecasting}

 \author{Yunshan Ma}
 \authornote{Equal contribution.}
 \affiliation{
     \institution{National University of Singapore}
 }
 \email{yunshan.ma@u.nus.edu}
 
 \author{Yujuan Ding}
 \authornotemark[1]
 \affiliation{
     \institution{The Hong Kong Polytechnic University}
 }
 \email{dingyujuan385@gmail.com}
 
 \author{Xun Yang}
 \authornote{Corresponding author.}
 \affiliation{
     \institution{National University of Singapore}
 }
 \email{xunyang@nus.edu.sg}

 \author{Lizi Liao}
 \affiliation{
     \institution{National University of Singapore}
 }
 \email{liaolizi.llz@gmail.com}

 \author{Wai Keung Wong}
 \affiliation{
    \institution{The Hong Kong Polytechnic University}
 }
 \email{calvin.wong@polyu.edu.hk}

 \author{Tat-Seng Chua}
 \affiliation{
     \institution{National University of Singapore}
 }
 \email{dcscts@nus.edu.sg}

\begin{abstract}
Fashion trend forecasting is a crucial task for both academia and industry. Although some efforts have been devoted to tackling this challenging task, they only studied limited fashion elements with highly seasonal or simple patterns, which could hardly reveal the real fashion trends. Towards insightful fashion trend forecasting, this work focuses on investigating fine-grained fashion element trends for specific user groups. 
We first contribute a large-scale fashion trend dataset (FIT) collected from Instagram with extracted time series fashion element records and user information.
Furthermore, to effectively model the time series data of fashion elements with rather complex patterns, we propose a Knowledge Enhanced Recurrent Network model (KERN) which takes advantage of the capability of deep recurrent neural networks in modeling time-series data. Moreover, it leverages internal and external knowledge in fashion domain that affects the time-series patterns of fashion element trends. Such incorporation of domain knowledge further enhances the deep learning model in capturing the patterns of specific fashion elements and predicting the future trends. 
Extensive experiments demonstrate that the proposed KERN model can effectively capture the complicated patterns of objective fashion elements, therefore making preferable fashion trend forecast.

\end{abstract}



\maketitle

\section{Introduction}
\textit{Karl Lagerfeld}~\footnote{https://wikipedia.org/wiki/Karl\_Lagerfeld} used to say that the essence of fashion is changeability. Fashion trend forecasting, aiming to master such change, is therefore of great significance in fashion industry. It enables fashion companies to develop products and establish marketing strategies more wisely. It also helps fashion consumers make better choices. Traditionally, to predict fashion trends, the staffs of forecasting companies travel across the world to observe the art, music, and other cultural factors that may influence fashion industry. Also, the staffs collect information of consumers' ways of living, thinking, and behaving~\cite{kim2013fashion}. 
However, the existing solutions mainly rely on subjective inferences of these forecasters, which may be less reliable and have large variations.

\begin{figure}[!t]
	\centering
	\includegraphics[scale = 0.3]{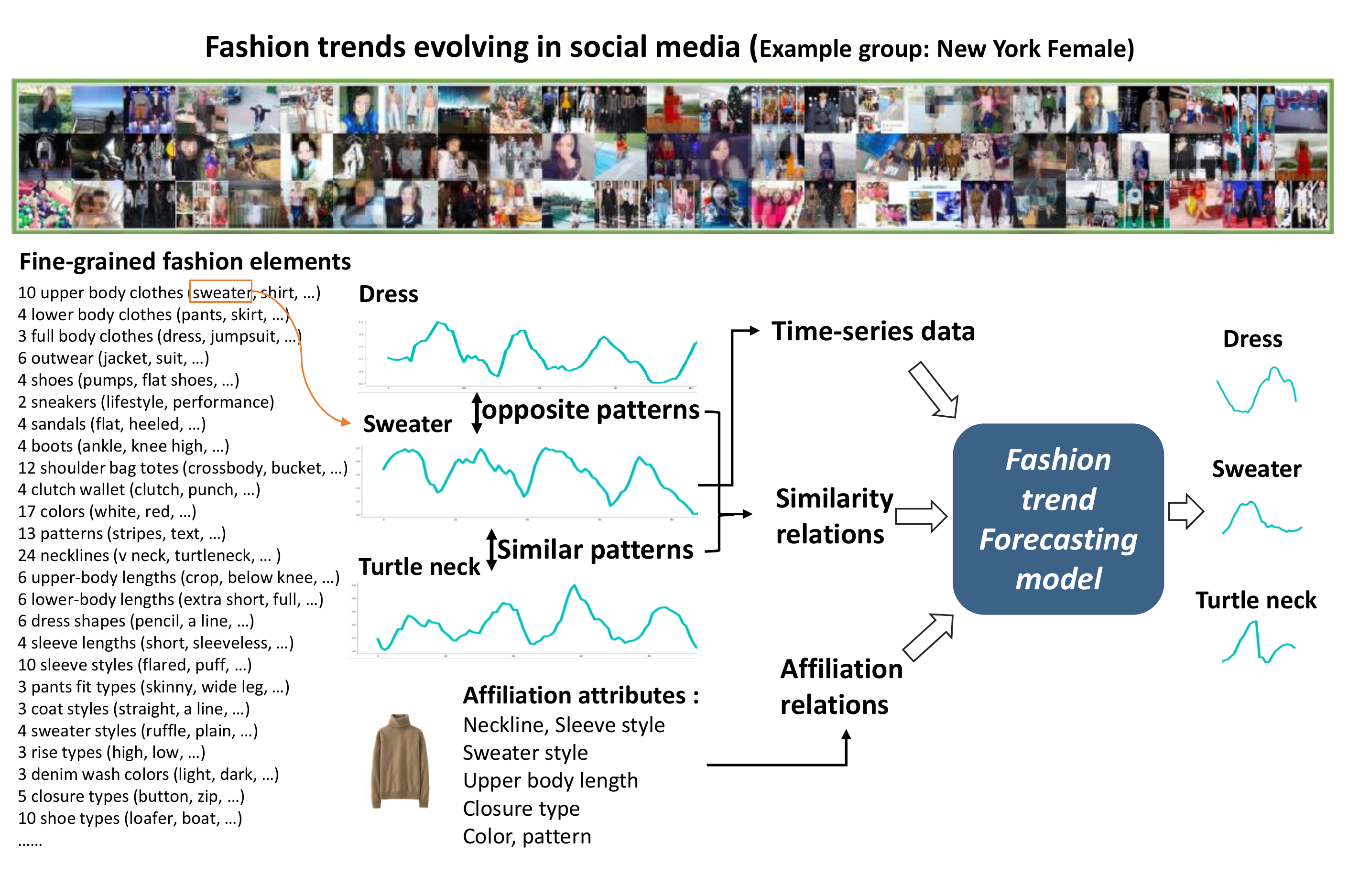}
	\vspace{-0.25in}
	\caption{The fashion trend forecasting task aims to predict the future trends of meaningful fashion elements.
	}
	\label{Fig:task}
	\vspace{-0.15in}
\end{figure}

In the recent decade, technological innovations such as Internet has accelerated the rate of fashion change, which makes fashion trend forecasting even more difficult. 
On the other hand, the advent of digital age has facilitated the accumulation of huge amounts of fashion-related data, which provides an alternative data-driven way of addressing the fashion trend forecasting task~\cite{al2017fashion}. 
This paper aims to mine useful fashion information from big historical data and predict the possible development of fashion for the future~\cite{Hidayati2014what,al2017fashion,mall2019geostyle}. There are two main research challenges for this task: 1) What kind of data should be used and analyzed in order to make meaningful and relevant fashion trend forecasting? 2) How to effectively model relevant data to make accurate predictions?

For the first challenge, the source data should contain abundant time series fashion information, and should also be of considerable scale to cover a rather long time period in order to reflect the evolution of fashion over time. Compared to e-commence or fashion show~\cite{vittayakorn2015runway,al2017fashion}, social media is a more appropriate data source because it sensitively and extensively records the fashion development with massive uploaded fashion-related images and comments everyday from multiple sources of end users, fashion bloggers and brands, \textit{etc}. Besides, rich information for both users and fashion items can be extracted from the images, meta data and other source data by the well-developed computer vision or other machine learning techniques. Although there exists datasets based on social media~\cite{mall2019geostyle}, they contain very limited fashion elements and are far from enough for forecasting meaningful and applicable fashion trends. Also, the information of users (such as age, gender or living location) that actually convey most fashion-related data is essential in fashion trend observation. Such user information, however, is neglected in existing datasets. Considering the limitations of existing datasets, in this paper, we build a new dataset with extensive fine-grained fashion elements, including category, attribute and style. It also covers a longer time period with richer user information. More details are introduced in Section 3.

For the second challenge, in order to make accurate data-driven fashion trend forecasting, the underlying patterns in the time series data need to be effectively captured. Though traditional models such as statistical models or matrix factorization have been effectively applied to model simple time series data~\cite{mall2019geostyle,matzen2017streetstyle}, they fall short of ability to make sound predictions for more complicated fashion trends. Recent advances of deep learning have provided great solutions for many tasks~\cite{lecun2015deep}. In particular, the recurrent neural networks (RNN) have demonstrated its superiority in modeling time series data and addressing relevant problems~\cite{cinar2017position,wen2017multi,fan2019multi}. However, such approaches have not been employed in the area of fashion trend analysis yet. 
On the other hand, most existing works predict the trend of each fashion element independently. However, according to common sense, fashion elements are not independent but well-correlated with each other in various ways. For example as shown in Figure~\ref{Fig:task}, the trend of \textit{sweater} shows similar pattern with that of \textit{turtle neck}, but nearly opposite with that of \textit{dress}. If we try to predict the trend of \textit{sweater}, we can apply the prediction results on both the \textit{turtle neck} and \textit{dress} to refine the prediction of \textit{sweater} based on their observed correlations. Furthermore, in fashion domain, there naturally exist taxonomic relations between elements, \textit{e.g.}, the affiliation relations between \textit{sweater} and all its affiliation attributes as shown in Figure~\ref{Fig:task}. Such taxonomic relations would result in relations among fashion trend patterns, which we should take advantage of in fashion trend modeling. In short, these types of prior domain knowledge describing the relations among fashion trends are non-trivial to model but helpful.

Driven by the above motivations, this paper presents a novel approach named Knowledge Enhanced Recurrent Network (KERN) for forecasting fashion trends of people in various groups. The proposed approach effectively models the time series data of fashion elements with rather complex patterns by using the Long-Short Term Memory (LSTM) encoder-decoder framework. More importantly, it incorporates two types of knowledge: internal and external knowledge. Specifically, for internal knowledge, it leverages the similarity relations of time series within dataset and introduces a triplet regularization loss based on pattern similarities. For external knowledge, it takes advantage of the affiliation relations of fashion elements within the taxonomy, and incorporates them by updating the embedding of fashion elements via message passing. The proposed KERN model incorporates both the time series information of single fashion element and the connectivity between this element and all related ones. We also exploit the user information for better modeling the different fashion trends for different groups of users by applying the semantic group representation.

In summary, the contributions of this work are three folds: 1) towards meaningful fashion trend forecasting, we contribute a large-scale fashion trend dataset based on Instagram, termed Fashion Instagram Trending (FIT); 2) to make sound trend forecasting, we propose a novel knowledge enhanced LSTM-based model (KERN) to effectively model the time series fashion trend data; and 3) we conduct extensive experiments and analysis of fashion trend forecasting on our FIT dataset and the GeoStyle dataset~\cite{mall2019geostyle}, and show that our KERN model is capable of capturing patterns in time series fashion trends data and effectively forecasting fashion trends.

\section{Related work} \label{related_work}

\subsection{Fashion Trend Analysis}
Recently, various tasks in fashion domain has attracted research interests such as fashion recognition~\cite{liu2016deepfashion,wang2018attentive,ma2019and}, retrieval~\cite{liu2012street,huang2015cross,liao2018interpretable,ding2019fashionhashing}, mix-and-match~\cite{song2017neurostylist,song2018neural,han2017learning,yang2019transnfcm, yang2019interpretable}, and visual try-on~\cite{han2018viton,hsieh2019fashionon}, \textit{etc}. Meanwhile, fashion trend analysis and forecasting has always been a classic research topic in fashion domain due to its significance in guiding the whole fashion industry. 
Hidayati \textit{et al.}~\cite{Hidayati2014what} analyzed fashion trends by proposing a framework to automatically detect fashion patterns which frequently occur within a fashion week. 
Vittayakorn \textit{et al.}~\cite{vittayakorn2015runway} extended such task to a larger dataset and studied both runway and real world fashion to produce quantitative analysis for fashion and trends. 
The main purpose of the above works is to analyze the current fashion styles, but not to make predictions. Likewise, Matzen \textit{et al.}~\cite{matzen2017streetstyle} studied clothing trends by statistical analysis, but based on a large-scale social media image dataset. Al-Halah \textit{et al.}~\cite{al2017fashion} proposed a fashion trends prediction model that uses nonnegative matrix factorization (NMF) to discover fashion styles and an exponential smoothing model to forecast the future of a style. However, the fashion styles they discovered are not real fashion style, but the cluster of certain kind of clothes with similar visual appearance. Comparatively, Mall \textit{et al.}~\cite{mall2019geostyle} explored very specific fashion elements and tried to find detailed fashion trends. They modeled the fashion trend signals of each target element with a basic combination of linear and cyclical components, which, according to the paper, were capable of capturing both coarse-level trends and fine-scale spikes. However, the limitation is that they only targeted limited fashion elements which showed simple patterns in their trend signals (such as with hat or not) and did not really include fine-grained fashion elements. In summary, existing works are still limited to statistical analysis or predicting trends of specific fashion styles or elements with simple patterns. More meaningful fashion elements involving complex trends are still yet to be explored.

\subsection{Time Series Forecasting} 
Fashion trend forecasting is also closely related to the time series forecasting problem which aims to predict the future based on the historical observations. Statistic models are classic solutions for time series forecasting problems, including the most representative autoregrassive (AR)~\cite{walker1931periodicity}, moving averages (MA)~\cite{slutzky1937summation}, improved autoregressive integrated moving average (ARIMA)~\cite{box1968some}, and others~\cite{winters1960forecasting,holt2004forecasting}. These models were found to be quite effective for forecasting structural data with high seasonality or simple trend. However, the real-life times-series signals are usually highly volatile and very difficult to model by these traditional methods. Recently, with the success of deep neural networks in a wide range of tasks, RNN, especially its variant LSTM~\cite{hochreiter1997long}, has shown its superiority in modeling sequential data and achieved superior performance in various applications of NLP~\cite{mikolov2010recurrent}, speech recognition~\cite{graves2013speech}, and also time series forecasting~\cite{cinar2017position,sutskever2011generating,langkvist2014review}.

Since fashion trend forecasting is a rather domain-specific task, leveraging abundant fashion knowledge in the forecasting task is a viable approach. Actually, exploiting domain knowledge, or external knowledge, to enhance the performance of deep learning models has achieved promising results lately in many tasks~\cite{wang2019explainable, liao2018knowledge, hu2018relation,jiang2018hybrid,venugopalan2016improving}.
Specifically, in time series forecasting problems like the stock price prediction, Feng \textit{et al.} found that incorporating domain knowledge of stocks (\textit{e.g}., companies within the same industry sector) can effectively help stock price forecasting \cite{feng2019temporal}. 
Despite of many successful applications, domain knowledge has not been well exploited in fashion trend forecasting.

\section{Problem Formulation and Dataset}
This paper focuses on the fashion trend forecasting problem, which aims to make prediction of future popularity with regard to each fashion element (\textit{e.g.}, \textit{white}, \textit{dress}, \textit{off-shoulder}, \textit{etc.}) for each user group (\textit{e.g.}, \textit{London female of age between 18 and 25}). Given a fashion element $f\in \mathcal{F}$ and a user group $g\in \mathcal{G}$, the temporal popularity of $f$ for $g$ is defined as a time series denoted as $\pmb{y}_g^f=(y_1, \cdots, y_t, \cdots)$, where $\mathcal{F}$ is the set of all fashion elements; and $\mathcal{G}$ is the set of all user groups. The value of the time series at each time step $t$ is defined as $y_t=N_t^{g,f}/N_t^{g}$, where $N_t^{g,f}$ is the number of the fashion elements $f$ at time point $t$ for group $g$; $N_t^{g}$ is the number of all fashion items (\textit{e.g, clothing, bags, shoes, and \textit{etc.}}) observed at time point $t$ for group $g$. Given the historical inputs within the time span of $[1, T]$, our aim is to forecast the future values of time $[T+1, T+T']$, where $T$ is the historical sequence length or time span, and $T'$ is termed as the forecast horizon (the number of steps ahead to forecast, $T'>1$). 

\begin{figure}[!t]
	\centering
	\includegraphics[scale = 0.42]{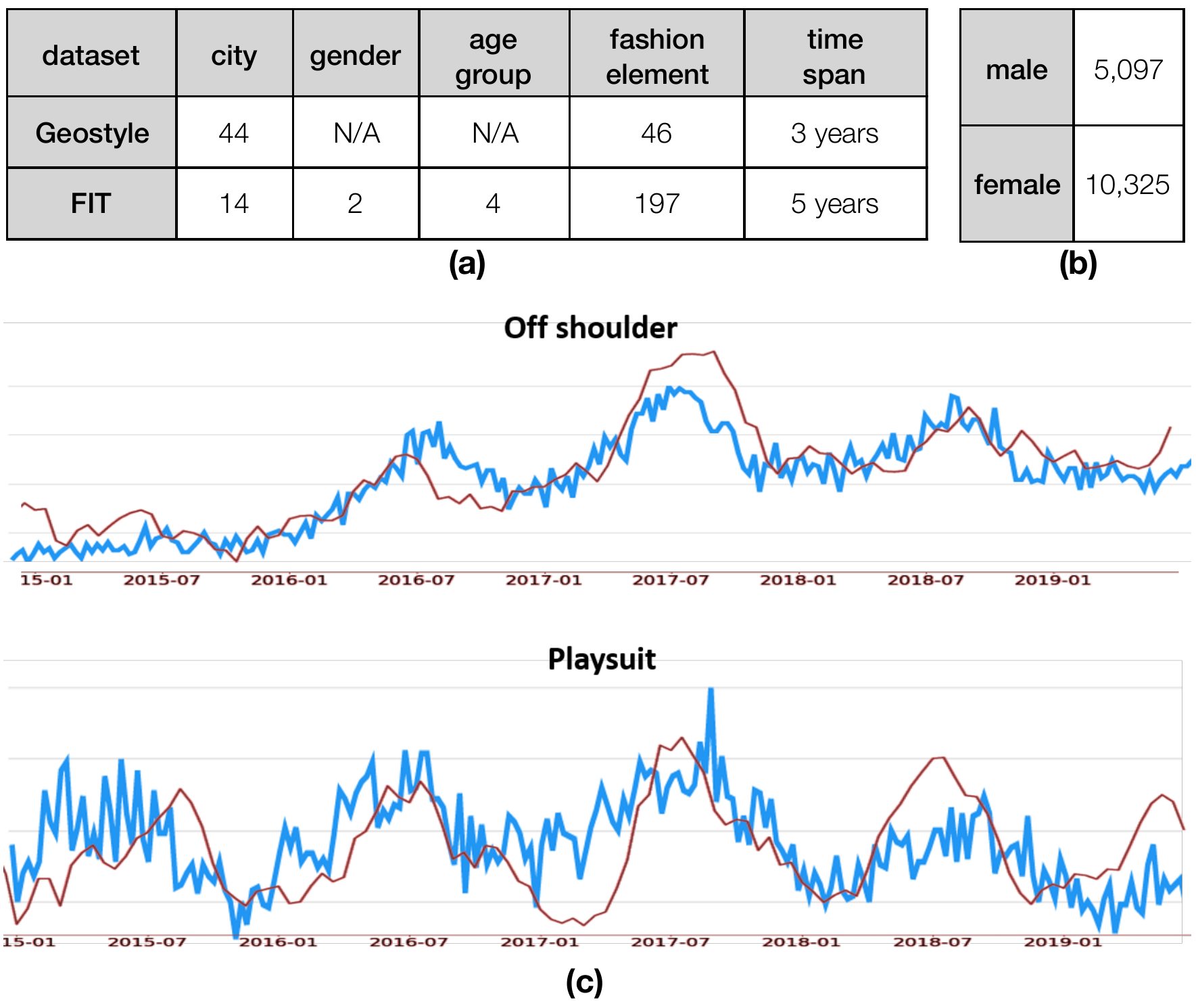}	
	\vspace{-0.05in}
	\caption{(a) (b) Statistics of the FIT dataset. (c) Two examples of the FIT dataset, where RED curves are from the FIT dataset and BLUE curves are from Google Trends (both examples belong to the group [New York, Female]).}
	\label{Fig:dataset}
	\vspace{-0.15in}
\end{figure}

Since none of the existing datasets supports this type of study, we contribute a new dataset based on the popular social media platform Instagram~\footnote{instagram.com}, termed as Fashion Instagram Trends (FIT). Figure~\ref{Fig:dataset} (a) and (b) show the statistical comparison between FIT and the Geostyle~\cite{mall2019geostyle}. It shows that the FIT dataset has more user information, richer fashion elements, and longer time span. 

Specifically, we crawl millions of posts uploaded by users from all over the world. To ensure quality of the crawled data, automated and manual filtering are conducted on the collected data, similar to that done in~\cite{ma2019and, ma2019automatic}. First, we leverage the pre-trained object detection model to detect person body~\cite{redmon2018yolov3} and face~\cite{zhang2016joint}. Images without face or body, or with abnormal-sized face or body are filtered out. Then, we drop posts with people that are not the corresponding account owner. Finally, we keep about \textit{680K} images in total. The annotation of the dataset is from two aspects: users and fashion elements, which will be introduced in detail below. 

For users, we collect three types of user information (\textit{i.e.}, age, gender and location), and then separate users into different groups based on the information. For each user we first apply the off-the-shelf age and gender detector tools~\cite{rothe2018deep,antipov2017effective,panis2016overview} on all of the users' posts (images), and then choose the dominant gender and the average age as the final gender and age. Posts detected as the opposite gender and with age differing from the detected age by over five years are dropped. We categorize the age of each user into four groups, that is: 0 to 18, 18 to 25, 25 to 40, and above 40. Next, we obtain the location based on the longitude and latitude data that comes with the post, and choose the most frequent one as the location of the user. Finally we keep a location set with 14 main cities across the world. The combination of the three types of user attributes forms a group, resulting in 74 groups. 

\begin{figure*}[tp]
	\centering
	\includegraphics[scale=0.46]{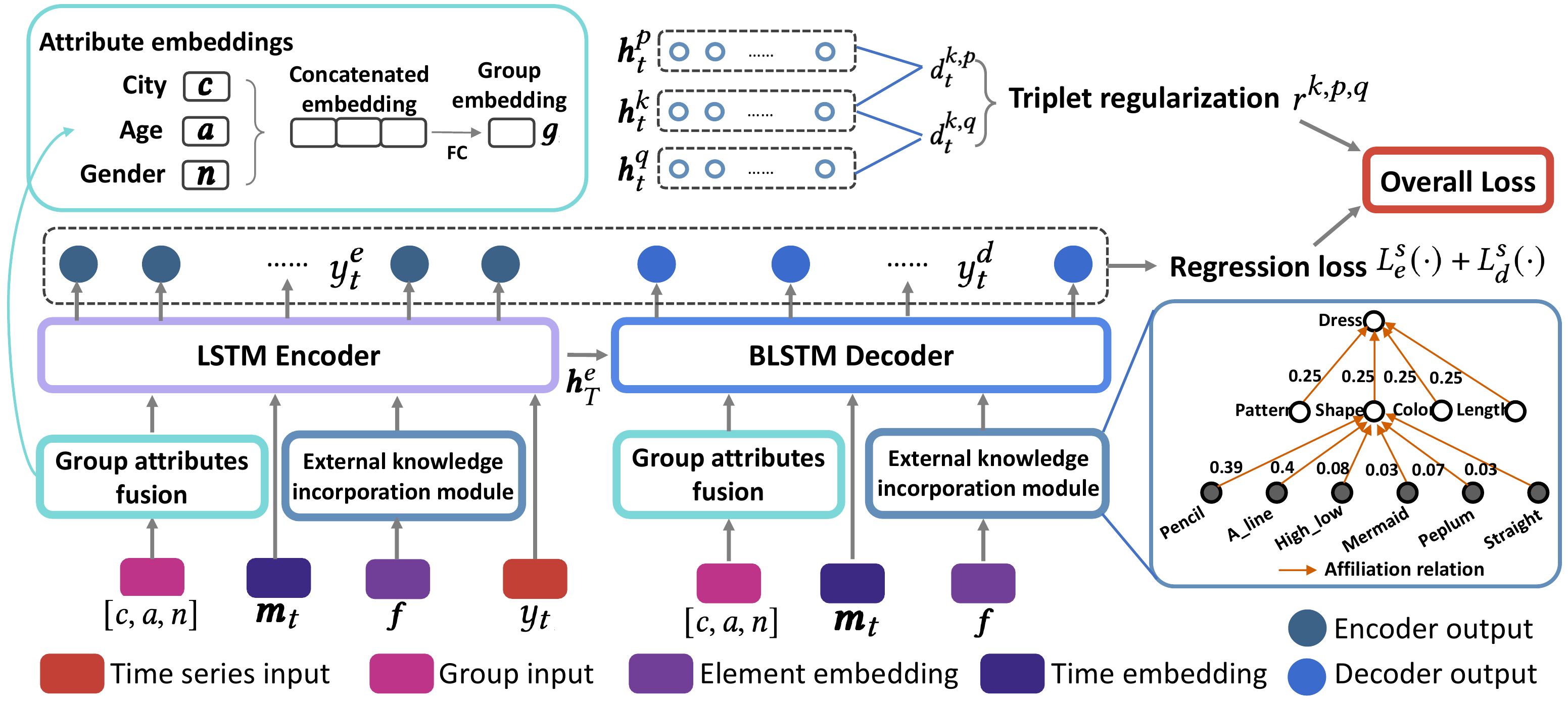}
	\vspace{-0.15in}
	\caption{The \textbf{K}nowledge \textbf{E}nhanced \textbf{R}ecurrent \textbf{N}etwork (KERN) framework. We utilize a basic LSTM encoder decoder framework to do multi-horizon trend forecasting and incorporate both internal and external knowledge via a triplet regularization term and a message passing module respectively.}
	\vspace{-0.15in}
	\label{Fig:overall_framework}
\end{figure*}

For fashion elements, we apply a commercial fashion tagging tool~\footnote{visenze.com} and extract three types of fashion elements (category, attribute and style) from the images, resulting in a total 197 of different fashion elements for the whole dataset. Each image is labelled with user group, time, and fashion elements after the annotation. We then calculate the popularity of each fashion element for each user group for every half month, resulting in a time series data. The post time of FIT dataset ranges from July 2014 to June 2019, spanning five years, which means that each time series has 120 data points. We further drop sparse time series with over \textit{50\%} of time points with no data. Finally, we obtain around \textit{8000} time series in total. Note that as our tags come from an existing tagging tool, which might contain some noise and result in a small bias of real fashion trends. However, we manually check part of the recognition results, and find that the average accuracy is relatively satisfying. More importantly, each time series data is a statistical ensemble of a group of users' data, thus the noise of each user on the final time series is hugely weakened. Besides, we comprehensively analyze fashion trends in FIT and compare them with that from \textbf{Google Trends}~\footnote{trends.google.com}, and observe highly similar patterns, which further validate the credibility of our FIT dataset (see examples in Figure~\ref{Fig:dataset}).

\section{Approach}
This paper aims to develop an end-to-end model to forecast the fashion trends given the historical inputs. 
First, we adopt the basic LSTM encoder decoder framework, which is able to incorporate both time series inputs and the associated sequence information into a unified model and make multi-horizon forecasting. Second, we add a triplet regularization term to explicitly incorporate the internal knowledge. Moreover, we introduce a message passing module to leverage the external knowledge extracted from the taxonomy. We name our proposed framework as \textbf{K}nowledge \textbf{E}nhanced \textbf{R}ecurrent \textbf{N}etwork (KERN), as shown in Figure~\ref{Fig:overall_framework}.

\subsection{Basic LSTM Encoder Decoder Framework}
Given a time series $(y_1, \cdots, y_T)$ indicating the past trend of fashion element $f$ for group $g$ within time period $[1,T]$, we aim to forecast the future values of the trend $(y_{T+1}, \cdots, y_{T+T'})$. The group $g$ is defined by the combination of three attributes: the city $c$, the age $a$ and the gender $n$, where $c \in \mathcal{C}$, $a \in \mathcal{A}$ ($\mathcal{C}$, $\mathcal{A}$ denote all cities and all age groups) and $n \in \{male, female\}$. We adopt a LSTM encoder decoder framework, including two main components: sequence feature embedding and LSTM encoder decoder network.
\subsubsection{Sequence Feature Embedding} \label{embedding}
Each time sequence $s$ is characterized by a group $g_s=[c_s, a_s, n_s]$ and a fashion element $f_s$ (we omit the subscription $s$ if there is no confusion thereafter). All the categorical features are converted into dense vector representations. Particularly, to get the group representation, we first convert the group features $c$, $a$, and $n$ into their embeddings $\pmb{c}\in \mathbb{R}^D$, $\pmb{a} \in \mathbb{R}^D$, and $\pmb{n} \in \mathbb{R}^D$ separately, where $D$ is the dimensionality of sequence feature embedding. We then adopt a linear layer to aggregate the three embeddings into one unified group representation:
\begin{equation}
    \pmb{g} = \pmb{W}_g[\pmb{c}, \pmb{a}, \pmb{n}] + \pmb{b}_g
\end{equation}
where $\pmb{W}_g \in \mathbb{R}^{3D\times D}$, $\pmb{b}_g \in \mathbb{R}^D$, and $\pmb{g}\in \mathbb{R}^D$. For each fashion element $f$, we directly covert it into a vector $\pmb{f} \in \mathbb{R}^D$.

\subsubsection{LSTM Encoder Decoder Network}
Most of the current methods for fashion trend forecasting~\cite{mall2019geostyle,al2017fashion} model each time series independently, overlooking the correlations among them. However, many fashion elements or groups have high correlations with each other, and the correlations can help to learn the trend patterns. For example, the seasonal trends of \textit{sweater} and \textit{t-shirt} are opposite with each other, if we learn one trend well and the other one will also perform well with high probability. Therefore, in this paper, we utilize a deep learning model: LSTM encoder decoder framework, by designing one model for all the time series instead of one for each, to implicitly capture the correlations among time series.

The LSTM encoder decoder framework consists of two parts: encoder and decoder, as shown in Figure~\ref{Fig:overall_framework}. The encoder is a LSTM network, which aims to map the historical inputs to latent representations $\pmb{h}_T^e$. Specifically, we concatenate the group representation $\pmb{g}$, the fashion element representation $\pmb{f}$, the timestep feature $\pmb{m}_t$ (the position of each point within one year, converted to vector representation thus $\pmb{m}_t\in \mathbb{R}^D$), and the trend value $y_t$ as the input of the encoder network at timestep $t$:
\begin{equation}
    \pmb{v}_t^e = [\pmb{g}, \pmb{f}, \pmb{m}_t, y_t]
\end{equation}
where $\pmb{v}_t^e \in \mathbb{R}^{3D+1}$. The output of the encoder LSTM is the hidden representations for the input sequence at timestep $t$, denoted as:
\begin{equation}
    \pmb{h}_t^e=LSTM^e(\pmb{v}_t^e;\pmb{h}_{t-1}^e)
\end{equation}
where $\pmb{h}_{t-1}^e, \pmb{h}_t^e \in \mathbb{R}^H$, and $H$ is the size of the hidden state. $\pmb{h}_{t-1}^e$ is the encoder hidden state one step ahead of $\pmb{h}_t^e$, 

The decoder network is a bi-directional LSTM, of which the initial hidden state is $\pmb{h}_T^e$, \textit{i.e.}, the last hidden state of encoder, and at each decoding step it takes the input feature and outputs the trend forecasting value. The input feature of decoder network at timestep $t$ is: $\pmb{v}_t^d = [\pmb{g}, \pmb{f}, \pmb{m}_t]$, which is different from $\pmb{v}_t^e$ by removing the trend value $y_t$ and thus $\pmb{v}_t^d \in \mathbb{R}^{3D}$. The bi-directional LSTM can propagate information from both forward and backward directions. Formally, we denote the hidden state from forward LSTM as $\overrightarrow{\pmb{h}_t^d}$ and from backward as $\overleftarrow{\pmb{h}_t^d}$. We can obtain the final hidden state $\pmb{h}_t^d$ by concatenating them as follows:
\begin{equation}
\left\{
\begin{aligned}
\overrightarrow{\pmb{h}_t^d} &= \overrightarrow{LSTM^d}(\pmb{v}_t^d; \overrightarrow{\pmb{h}_{t-1}^d}) \\
\overleftarrow{\pmb{h}_t^d} &= \overleftarrow{LSTM^d}(\pmb{v}_t^d; 
\overleftarrow{\pmb{h}_{t+1}^d}) \\
\pmb{h}_t^d &= [\overrightarrow{\pmb{h}_t^d}, \overleftarrow{\pmb{h}_t^d}]
\end{aligned}
\right.
\end{equation}
where $\overrightarrow{\pmb{h}_t^d}, \overrightarrow{\pmb{h}_{t-1}^d}, \overleftarrow{\pmb{h}_t^d}, \overleftarrow{\pmb{h}_{t+1}^d} \in \mathbb{R}^H$, and $\pmb{h}_t^d \in \mathbb{R}^{2H}$. The prediction is made based on the hidden state of each step. Due to the teacher mechanism leveraged in the encoder network, the predictions are made in both the encoder and decoder stages during training. But for testing, predictions only happen in the decoder stage. Particularly, we apply linear layers for making predictions for encoder and decoder respectively:
\begin{equation}
\left\{
\begin{aligned}
y_t^e &= \pmb{W}_e\pmb{h}_t^e + b_e \\
y_t^d &= \pmb{W}_d\pmb{h}_t^d + b_d
\end{aligned}
\right.
\end{equation}
where $\pmb{W}_e, \pmb{W}_d \in \mathbb{R}^{1\times 2H}$ and $b_e, b_d \in \mathbb{R}$ are the parameters for the linear layer; $y_t^e, y_t^d \in \mathbb{R}$ are the forecasting value at each timestep for encoder and decoder respectively. We use L1 loss to train the LSTM, including the encoder loss $L_e(\cdot)$ and decoder loss $L_d(\cdot)$:
\begin{equation}
    L_{LSTM} = L_e(\pmb{y}_e, \pmb{y}_e^{\ast}, \pmb{\theta}_e) + L_d(\pmb{y}_d, \pmb{y}_d^{\ast}, \pmb{\theta}_d)
\end{equation}
where $\pmb{\theta}_e$, $\pmb{\theta}_d$ are the model parameters for encoder and decoder respectively; $\pmb{y}_e, \pmb{y}_e^{\ast} \in \mathbb{R}^{(T-1)}$ are the prediction and ground-truth of the encoder sequence; and $\pmb{y}_d, \pmb{y}_d^{\ast} \in \mathbb{R}^{T'}$ are the prediction and ground-truth of the decoder sequence. 

\subsection{Knowledge Incorporation}
Formally, we leverage two types of knowledge: internal knowledge within the dataset and external knowledge extracted from the fashion element taxonomy.
\subsubsection{Internal Knowledge}
Fashion trend sequences demonstrate high correlation with each other. Particularly, some sequences have similar patterns while some sequences have opposite patterns. For example, as shown in Figure~\ref{Fig:task}, the time series pattern of \textit{turtle neck} is similar with \textit{sweater} but opposite with \textit{dress}. Such prior knowledge is valuable to guide the learning of the model. We deem that the hidden representations of time series with similar patterns should be close to each other, and that with opposite patterns should be far away from each other. We generalize such \textit{similar-opposite} relation to general \textit{close-far} relation, that is: given one fashion trend sequence $\pmb{y}^k$, we can always find another two series $\pmb{y}^p$ and $\pmb{y}^q$, where $\pmb{y}^p$ is closer to $\pmb{y}^k$ than $\pmb{y}^q$, and $\pmb{y}^k, \pmb{y}^p, \pmb{y}^q \in \mathbb{R}^{(T+T')}$.We use L1 distance (denoted as $\lVert\cdot\rVert$ thereafter) to measure the similarity. Formally these three sequences should satisfy: \begin{equation}
    \lVert \pmb{y}_k - \pmb{y}_p \rVert < \lVert \pmb{y}_k - \pmb{y}_q \rVert
\end{equation}
We construct a triplet regularization term $r^{k,p,q}$ as follows: 
\begin{equation}
\left\{
\begin{aligned}
d_t^{k,p} &= \lVert \pmb{h}_t^k - \pmb{h}_t^p \rVert \\
d_t^{k,q} &= \lVert \pmb{h}_t^k - \pmb{h}_t^q \rVert \\
r^{k,p,q} &= \frac{1}{T+T'-1}\sum_{t=0}^{T+T'-1}{max(0, d_t^{k,p} - d_t^{k,q})}
\end{aligned}
\right.
\label{eq8}
\end{equation}
where $\pmb{h}_t^k$, $\pmb{h}_t^p$, and $\pmb{h}_t^q$ are the LSTM hidden states in both encoder and decoder stages for sequences $k$, $p$, $q$ respectively; $d_t^{k,p}, d_t^{k,q} \in \mathbb{R}$ are the hidden state distances between $(k,p)$ and $(k,q)$ respectively. We randomly sample the $(k,p,q)$ triplet from the whole dataset. Thus the final loss of our framework is as follows:
\begin{equation}
    L^{k,p,q} = \sum_{s\in \{k,p,q\}}{\left(L_e^s(\cdot) + L_d^s(\cdot)\right)} + \lambda r^{k,p,q}
\end{equation}
where $\lambda$ is a hyper parameter determining the weight of the regularization term.
\subsubsection{External Knowledge}
\begin{figure}[!htp]
	\centering
	\includegraphics[scale=0.6]{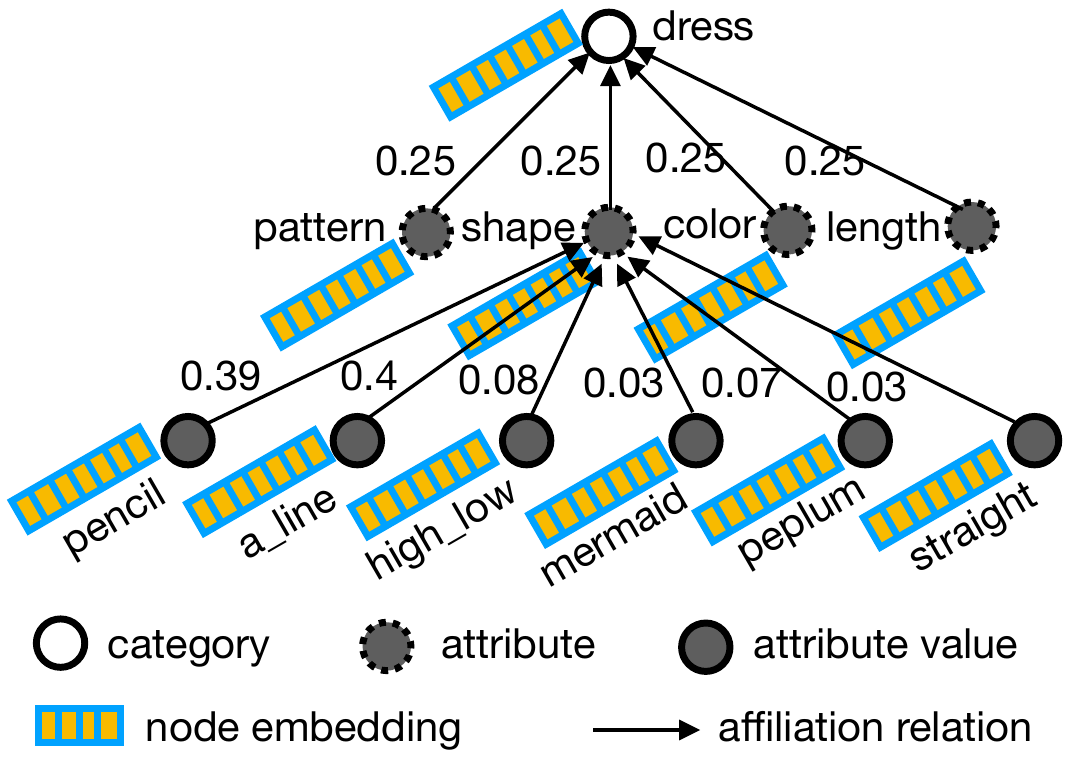}
	\vspace{-0.1in}
	\caption{An illustration of external knowledge incorporation module. 
	}
	\vspace{-0.1in}
	\label{Fig:kg_illustration}
\end{figure}

In addition to the internal knowledge of \textit{close-far} relations observed from the dataset, we leverage external knowledge from fashion element taxonomy to help in enhancing the model. Generally speaking, all the fashion elements are usually organized into a hierarchical taxonomy with a tree structure. There exist affiliation relations between children nodes and their associated parent nodes, which will further affect their corresponding trend sequences. For example, if we find that the trend of the attribute \textit{peplum} goes up, it is highly possible that the category \textit{dress} also goes up since \textit{peplum} is an attribute of category \textit{dress}. Different from the aforementioned similarity correlation which is pair-wise, the affiliation relation is many-to-one. For instance in Figure~\ref{Fig:kg_illustration}, the category \textit{dress} has four attributes and the attribute \textit{shape} has six values. To model such complicated relationships, we propose to construct a tree among all the fashion elements and conduct message passing between nodes with affiliation relationships.

A part of the constructed tree is shown in Figure~\ref{Fig:kg_illustration}. Basically, we have three types of nodes in this tree: \textit{category}, \textit{attribute}, and \textit{attribute value}, and the affiliation relations are between \textit{attribute} and \textit{category}, \textit{attribute value} and \textit{attribute}. As mentioned in Section \ref{embedding}, each fashion element $f$ is converted to a vector representation $\pmb{f}$. Therefore we conduct message passing among those embeddings, \textit{i.e.}, passing messages from children nodes to their parent nodes. The message passing for node $i$ is as follows: 
\begin{equation}
\left\{
\begin{aligned}
  \pmb{m}_i &= \sum_{j\in{N_i}}w_j\times{\pmb{f}_j} \\
  \pmb{f}_i &\leftarrow \pmb{f}_i + \pmb{m}_i
\end{aligned}
\right.
\end{equation}
where $\pmb{m}_i \in \mathbb{R}^D$ is the message passed from its children nodes, $N_j^s$ is the set of nodes that have affiliation relations with node $i$, and $w_j$ is the weight of each relation and can be learned during training. Note that we initialize the $w_j \in \mathbb{R}$ as the portion of node $j$ out of all nodes affiliated to the parent node, thus $\sum_{j\in{N_i}}{w_j} = 1$.

\begin{table*}[!htp]
  \caption{Performance of KERN and baselines for fashion trend Forecasting (the lower is better)}
  \vspace{-0.15in}
  \label{tab:1}
  \setlength{\tabcolsep}{6mm}{
  \begin{tabular}{ccc|cc|cc}
    \toprule
    Dataset &\multicolumn{2}{c|}{GeoStyle} &\multicolumn{4}{c}{FIT}  \\
    \midrule
    ~\multirow{2}*{\texttt{Method}} &\multicolumn{2}{c|}{Half year}  &\multicolumn{2}{c|}{Half year}  &\multicolumn{2}{c}{One year}\\
    ~ &MAE &MAPE  &MAE  &MAPE  &MAE  &MAPE   \\
    \midrule
    \texttt{Mean} &0.0292 &25.79  &0.132  &65.31  &0.135  &63.21   \\
    \texttt{Last} &0.0226 &21.04  &0.125  &46.45  &0.147  &54.04   \\
    \texttt{AR} &0.0211	&20.69	&0.114	&54.36	&0.119	&51.96 \\
    \texttt{VAR} &0.0150 &17.95	&0.157	&62.97	&0.126	&47.35 \\
    \texttt{ES} &0.0228	&20.59	&0.133	&55.29	&0.150	&57.42\\
    \texttt{Linear} &0.0365 &24.40 &0.112 &43.30 &0.133 &45.89 \\ 
    \texttt{Cyclic} &0.0165	&16.64	&0.129	&49.92	&0.143	&51.66\\
    \texttt{GeoStyle} &0.0149	&16.03	&0.136	&52.40	&0.149	&53.14  \\
    \textbf{KERN} &\textbf{0.0134}  &\textbf{14.24}  &\textbf{0.083}	&\textbf{30.02}	&\textbf{0.094}	&\textbf{33.45}  \\
    \midrule
    \midrule
    \textbf{improv(\%)} &\textbf{10.07} &\textbf{11.17}  &\textbf{25.89}	&\textbf{30.67}	&\textbf{21.01}	&\textbf{27.11}\\
    \bottomrule
  \end{tabular}}
  \vspace{-0.1in}
\end{table*}

\section{Experiments}
To verify the effectiveness of our proposed approach, we conduct extensive experiments on two datasets. 
In particular, we are interested in the following research questions: \\
(1)\textbf{RQ1}: Does our KERN model outperform current state-of-the-arts methods in predicting future fashion trend? \\
(2)\textbf{RQ2}: Whether the introduced relation modules help in improving the performance and how do they help? \\
(3)\textbf{RQ3}: How does the proposed model perform in trend forecasting in terms of specific fashion elements, and based on that, how can the model produce insightful fashion trend forecasting. 

\subsection{Experimental Settings}
\begin{figure}[!htp]
	\centering
	\includegraphics[scale = 0.36]{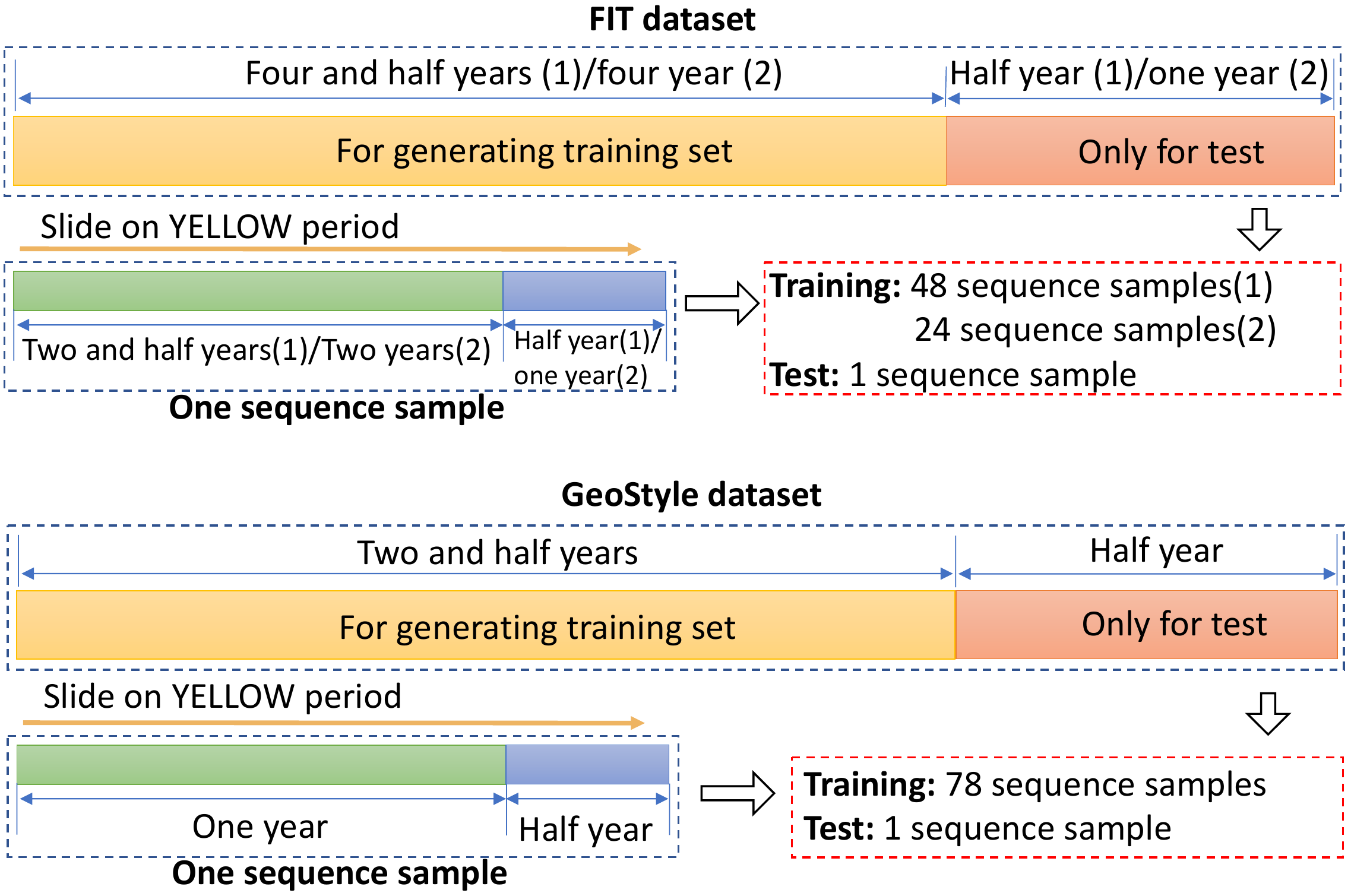}	\caption{Illustration of generating samples for each time series. Each sample is composed of the input sequence (GREEN) and output sequence (BLUE). Sliding window strategy is applied to generate the training samples. Note that there are two settings on FIT dataset, marked by (1) and (2).}
	\vspace{-10pt}
	\label{Fig:data_preparing}
\end{figure}

\textbf{Experimental Setup}. We apply two fashion trend forecasting datasets, our proposed FIT dataset and the GeoStyle dataset~\cite{mall2019geostyle}. 
For GeoStyle, we take one-year of data (52 data points) as input to predict the output of the following half year (26 data points). We design two settings on FIT, 1) use two years of data (48 data points) as input and predict the output of the following half-year (12 data points); and 2) use two years of data (48 data points) as input and predict the output of the following one-year (24 data points). Since Geostyle dataset only has one attribute (city) and does not have user attributes of age and gender, we simplify the fusion of group attributes and directly use the city embedding as group embedding. Sliding windows strategy was applied on both datasets to generate the training and testing samples. More details of data preparation can be found in Figure~\ref{Fig:data_preparing}, which shows the procedure of splitting each time series into shorter samples. We use Mean Absolute Error (MAE) and Mean Absolute Percentage Error (MAPE) as the evaluation metrics~\cite{mall2019geostyle}. \\
\textbf{Implementation Details}. We set the embedding size of user embeddings (including age embedding, gender embedding and city embedding), fashion element embedding and time embedding to 10 and the hidden size of both encoder and decoder LSTM network to 50. The hyper parameter $\lambda$ for regularization is set to $2\times10^{-4}$. During training, we randomly sample a batch of 400 different time series for each iteration. For each experimental setting, we train one KERN model for all fashion elements. For each testing sample, we calculate the performance of the odd data points as the validation results, and that of even data points as the testing results.\\
\textbf{Baseline Methods}. We select several state-of-the-arts time series prediction methods to compare with the proposed KERN model: \\
\textbf{Mean} and \textbf{Last}: They use the mean value or the value of last point of the input historical data as the forecasting value. \\
\textbf{Autoregression (AR)}: It is a linear regressor which uses a linear combination of last few observed values as the forecasting value.\\
\textbf{Vector Autoregression (VAR)}: VAR is a generalization of the AR by allowing for more than one evolving variable.\\
\textbf{Exponential Smoothing (ES)}~\cite{al2017fashion}: It aggregates all the historical values with an exponential decayed weight, the more recent values have higher impact on the future's forecast. \\
\textbf{Linear} and \textbf{Cyclic}~\cite{mall2019geostyle}: They are linear or cyclical parametric model which let historical values to fit the specific predefined model. \\
\textbf{Geostyle}~\cite{mall2019geostyle}: It is a parametric model combining a linear component and a cyclical component. It is the state-of-the-art fashion trend forecasting method on Geostyle dataset.
\begin{table}
  \caption{Contribution of different knowledge in KERN model (MAE results). `-E' means without external knowledge and `-I' means without internal knowledge and `-IE' means neither is used. Since there is no taxonomy among Geostyle's fashion elements, KERN-E is left empty.}
  \vspace{-0.15in}
  \label{tab:2}
  \setlength{\tabcolsep}{4.2mm}{
  \begin{tabular}{cccc}
    \toprule
    Dataset &GeoStyle &\multicolumn{2}{c}{FIT}  \\
    \midrule
    Prediction &Half year  &Half year  &One year\\
    \midrule
    \texttt{KERN-IE} &0.0137  &0.0840 &0.0966 \\
    \texttt{KERN-E} &-  &0.0835  &0.0953 \\
    \texttt{KERN-I}&0.0134 &0.0831  &0.0942 \\
    \texttt{KERN} &0.0134  &0.0836  &0.0939\\
    \bottomrule
  \end{tabular}}
  \vspace{-0.15in}
\end{table}

\begin{figure}[!htp]
	\centering
	\includegraphics[scale = 0.35]{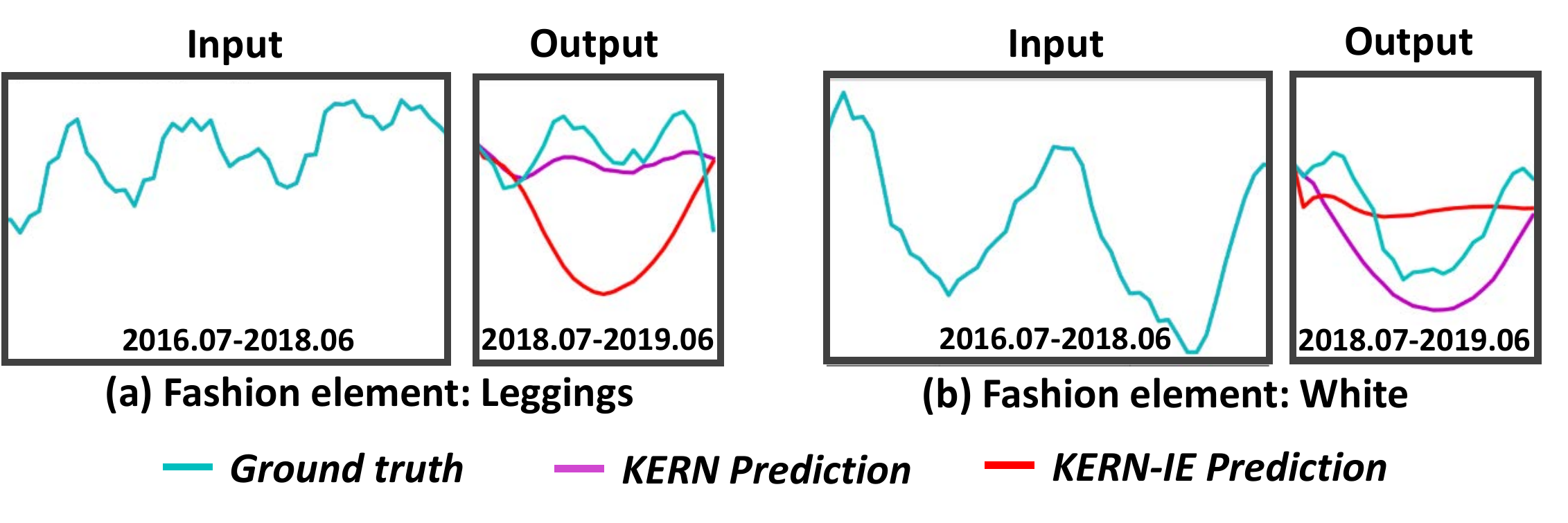}
	\vspace{-0.15in}
	\caption{Two examples of trends forecasting made by KERN and KERN-IE models.  
	}
	\label{Fig:ablation}
	\vspace{-0.15in}
\end{figure}

\subsection{Overall Performance on Fashion Trend forecasting(RQ1)}
We first evaluate our KERN model in terms of fashion trend forecasting by comparing its performance with several classic and state-of-the-arts baselines. The overall results are shown in Table~\ref{tab:1}. Based on the results, we have the following observations: 

(1) The proposed KERN model consistently yields the best performance on both datasets and under all experimental settings.  Specifically, on the GeoStyle, KERN is the only method to achieve the MAE result of lower than \textbf{0.014} and MAPE lower than \textbf{15}. On the two settings of FIT dataset, MAE and MAPE results of all baselines are over \textbf{0.11} and \textbf{40}, but our KERN model achieve MAE of under \textbf{0.1} and MAPE of around \textbf{30}, both show quite superior performance. 

(2) On the FIT dataset, the KERN method outperforms all baselines with large margins. As the FIT dataset contains much more fine-grained fashion elements, more user information, and more realistic and complex time series patterns, it is more challenging to model. Therefore, the baseline methods do not perform well. However, our KERN method is able to capture such complex patterns compared with other baselines as it better models nonlinearity in data by the LSTM encoder decoder framework and leverages the abundant domain knowledge. That also explains why our model achieves limited improvement on GeoStyle.
\begin{figure}[!t]
	\centering
	\includegraphics[scale = 0.28]{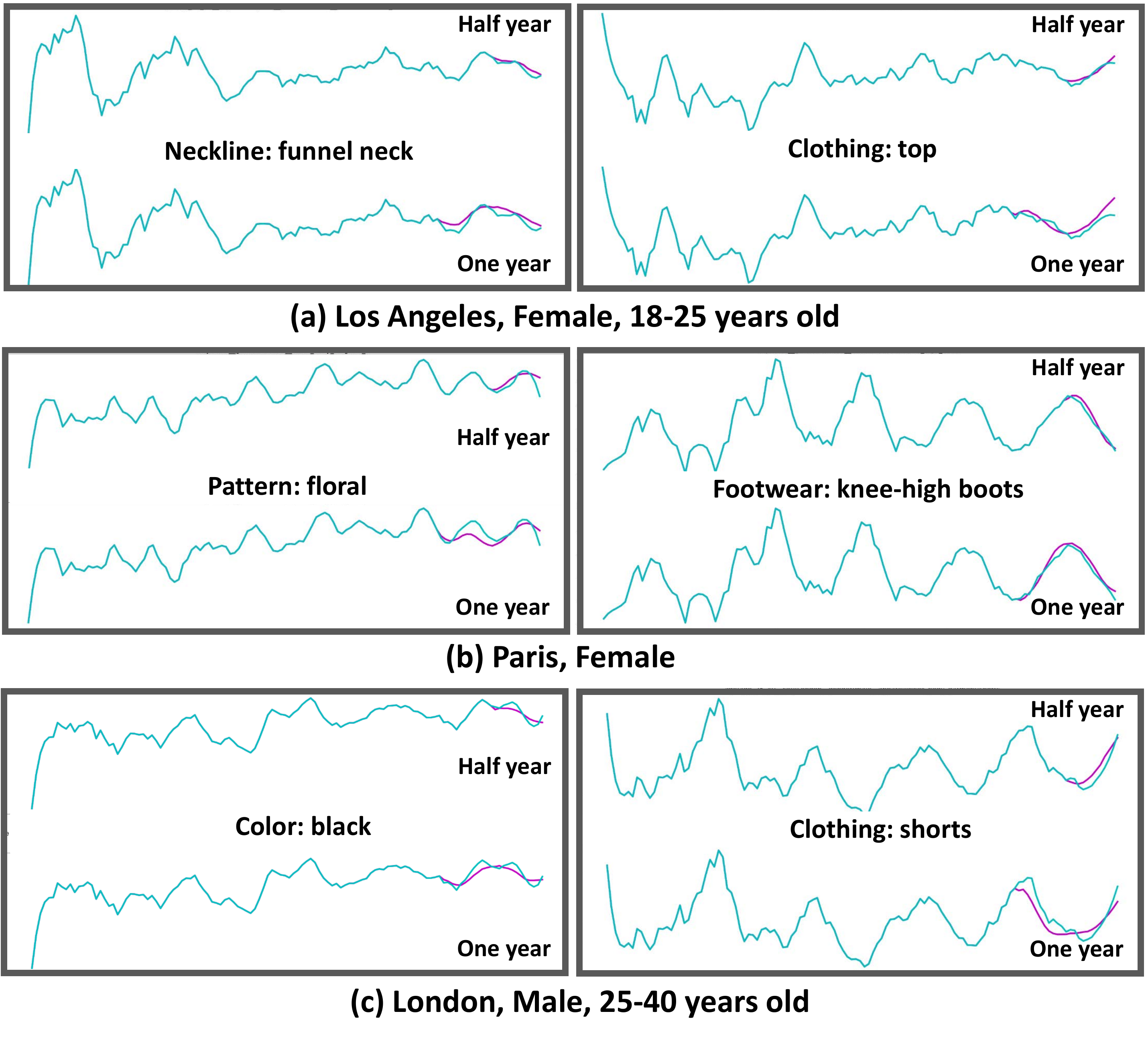}
	\vspace{-0.15in}
	\caption{Examples of trend forecasting for different fashion elements and user groups. GREEN curves are ground truth and PURPLE curves are prediction results by KERN.
	}
	\label{Fig:case1}
\end{figure}

\begin{figure}[!t]
	\centering
	\includegraphics[scale = 0.4]{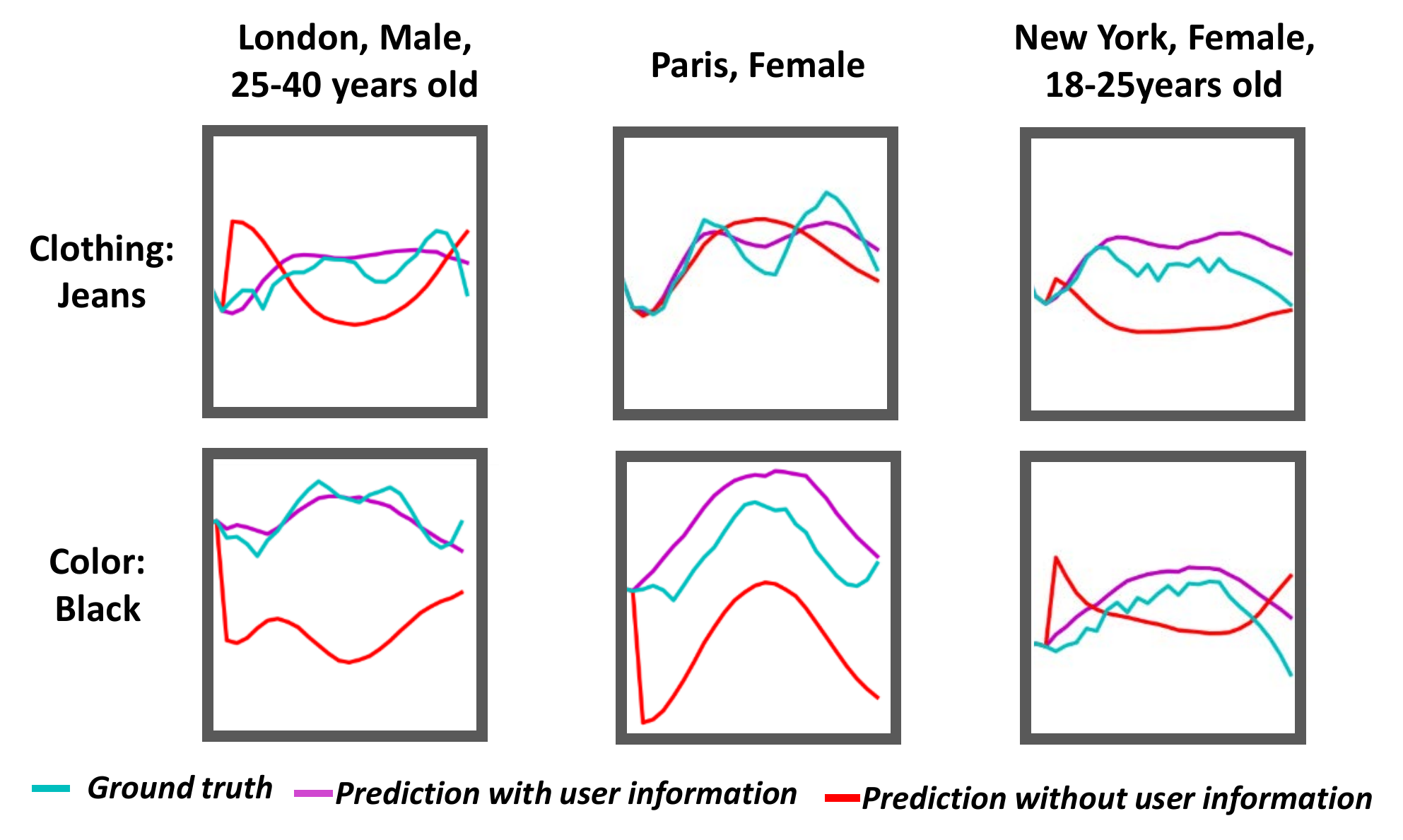}
	\vspace{-0.15in}
	\caption{The prediction results of one fashion element of different user groups. The performance is better when applying user information in data modeling.
	}
	\label{Fig:group}
	\vspace{-0.15in}
\end{figure}

\begin{figure*}[!t]
	\centering
	\includegraphics[scale = 0.53]{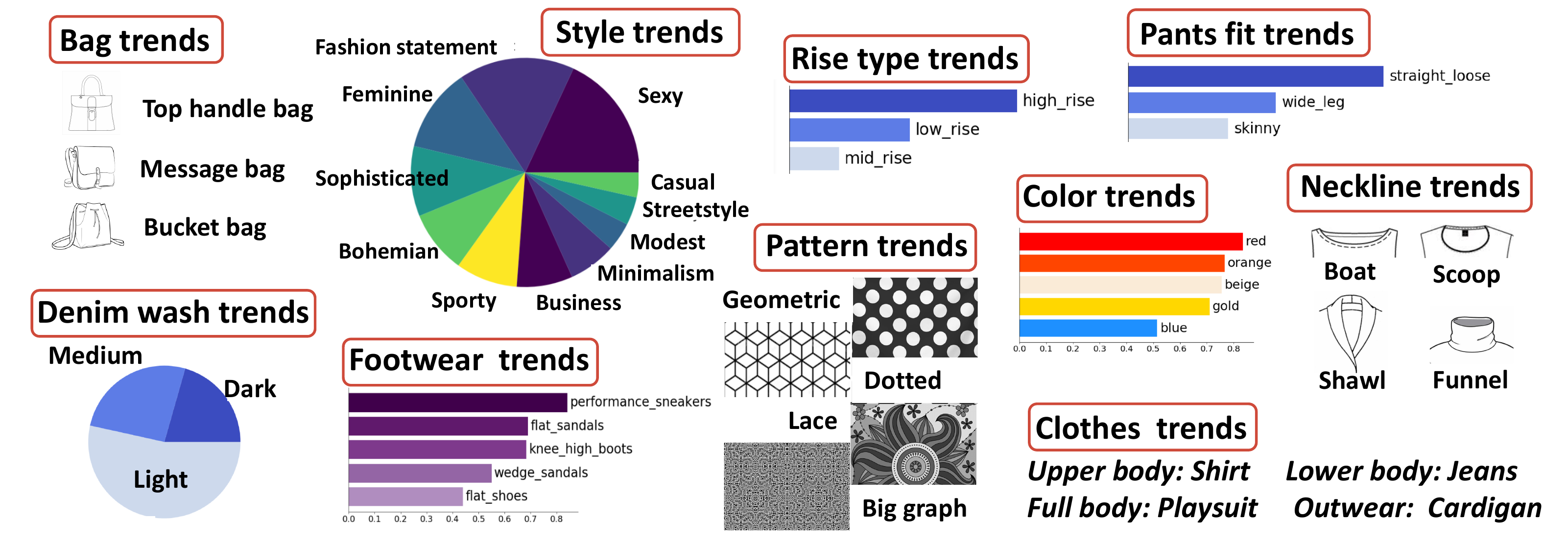}
	\vspace{-0.1in}
	\caption{The fashion trend report generated based on KERN forecasting results for 2018.07. This report is targeted on women in Paris, and predicted based on fashion records from 2016.07 to 2018.06.   
	}
	\label{Fig:report}
	\vspace{-0.1in}
\end{figure*}

(3) Most methods perform better in half-year prediction than in one-year prediction on the FIT dataset, including our KERN model. Such results are reasonable since first, the one-year prediction requires to forecast data with longer time horizon. Second, such setting reduces the quantity of training data (see Figure~\ref{Fig:data_preparing}).

\subsection{Discussion on Effectiveness of Knowledge Incorporation (RQ2)}
Here we conduct experiments and discuss the effectiveness of incorporating knowledge in the fashion trend forecasting model. In particular, two types of knowledge are exploited. The first is the internal knowledge of similarity relations, utilized by introducing triplet regularization term in the loss function. The second is the external knowledge of affiliation relations built from taxonomy, utilized by updating the fashion element embeddings by message passing according to the affiliation tree. 

To evaluate the effectiveness of each type of knowledge, we test the model by removing each type one at a time, as well as both simultaneously. Specifically, the model without using internal knowledge (KERN-I) ignores the triplet regularization loss (Eqn.~\ref{eq8}) while the model without using external knowledge (KERN-E) is not equipped with the external knowledge incorporation module. The KERN-IE model contains neither. From the performance reported in Table~\ref{tab:2}, we can see that removing either affiliation relations or similarity relations could degrade the performance of our model. The difference made by incoporating knowledge is more significant for longer forecasting on FIT dataset, which shows that leveraing knowledge could be particularly helpful for complicated and challenging data. Figure~\ref{Fig:ablation} shows two examples that compare the trend forecasting results of KERN and the ablated KERN-IE. The prediction results of KERN are clearly better than that of KERN-IE, which further shows that KERN benefits from effectively applying internal and external knowledge.


\subsection{Fashion Trend Analysis (RQ3)}
To further illustrate the effectiveness of our KERN model for fashion forecasting, we show more visualization results in this part. We first show the one-year prediction and half-year prediction of six representative fashion elements (including clothing category, footwear category, color, pattern and attributes) for the three user groups as shown in Figure~\ref{Fig:case1}. We can see that, in general, KERN can predict the trends of fashion element very well for both the one year or half year prediction, even for those with rather complex patterns. Results in Figure~\ref{Fig:group} show that the same fashion element between different user groups can be different. It also demonstrates that the KERN method can effectively predict the different trends by leveraging user information because the prediction results using user information is much better than those without. 

Based on the forecasting of various types of fashion elements, we can further generate a comprehensive and professional fashion trend report (see Figure~\ref{Fig:report}) that covers a large number of fashion trends ranging from category, color, pattern, style, or even detailed attributes such as denim wash colors.

\section{conclusion and future work}
This paper addresses the fashion trend forecasting problem based on social media, aiming to mine the complex patterns in the historical time-series records of fashion elements and accordingly predict the future trends. 
An effective model, Knowledge Enhanced Recurrent Network (KERN) is proposed to capture the complex patterns in the time-series data and forecast fashion trends.

Although much effort has been made and desirable results have been achieved, there are some aspects that can be further improved in the future. First, more user information should be explored such as occupations or hobbies. 
Second, multiple sources of knowledge should be considered such as the fashion analysis derived from professional fashion magazines, fashion bloggers and brands.

\section*{acknowledgement}
This research is supported by the National Research Foundation, Singapore under its International Research Centres in Singapore Funding Initiative, and The Hong Kong Polytechnic University (project code: RHQK). Any opinions, findings and conclusions or recommendations expressed in this material are those of the author(s) and do not reflect the views of National Research Foundation, Singapore. We also appreciate the fashion recognition API service provided by Visenze.

\newpage
\bibliographystyle{ACM-Reference-Format}
\balance
\bibliography{main}

\end{document}